\documentstyle[12pt,a4]{article}
\title{The Ising Like Statistical Models for Studying the Dynamics of the  
       Financial Stock Markets}
\author{Dorina Andru Vangheli, Gheorghe Ardelean\\ 
        West University of Timisoara, Romania}
\begin{document}
\sloppy
\maketitle
\begin{abstract}
In this paper, we present the possibility of using the Ising like models
to explain by Statistical Physics means the connection between the financial
discontinuities (herd behavior, bubbles, crashes) and "critical points"
in physical of phase transitions due to cooperative effects. For this purpose
there are investigated $\frac{1}{2}$ and $\frac{3}{2}$ Ising like spin models
which involve one or more macroscopical order parameters, which can have 
nonanlitical behavior in the critical domain.
\end{abstract}

\section{Q - M Dependence}

As it was performed earlier, [1], in econophysics there is a key assumption 
that a crash in financial market may be caused by local self-reinforcing imitation
between "noise traders" (traders who have no own opinion and decide depending on the "total opinion"). 
If the tendency for noise traders to imitate their nearest 
neighbors increases, up to a certain point, called the "critical point", all noise
traders may place the same order at the same time, thus causing a {\bf crash}.
At this point of view, it is clear that the market crashes are caused
by the slow buildup of long-range correlations, leading to a collapse of the stock
market in one critical state. It is the "criticality", which, in the physical word,
determines one particular state, in which some well-defined quantities present
an explosion to infinity. In this case there are present so called "cooperative phenomena".

For these reasons, as a microscopic modeling, it was considered [2] the simplest statistical
lattice model, i.e. the Ising -  $\frac{1}{2}$ spin model. 

The Ising - $\frac{1}{2}$ lattice-gas models  describe a large class of phase transitions
characterized by cooperative phenomena (such as, in physics:
the adsorption and absorption of a gas, the condensation of an fluid, the freezing
of an liquid, the order-disorder transition in a binary alloy, some magnetic 
and electric transitions, etc.). All these phase transitions are characterized, macroscopically, 
by a single-order parameter, which, depending on the considered case,  can  
be: density, concentration, magnetization, electric polarizability, etc.

For implementation of  this model in econophysics, it 
must be made the assumption that there is a network of agents, indexed as
$i=1,2,..,I$, so that each agent can be in one of the two possible "spin states"
$S_i \in \{ -1,+1\}$, interpreted as "buy", respectively, "sell", as in Ising - $\frac{1}{2}$

statistical model, characterized by a hamiltonian form 
\begin{equation}
{\cal H}=J\sum _{i>j}S_iS_j +H\sum _i S_i
\end{equation}
and having an ordering parameter, which, in context of the alignment of the spins,
create a macroscopic magnetization 
\begin{equation}
M=\frac{1}{I}\sum _i S_i
\end{equation}
In the absence of the noise traders global influence, there is realized the 
equilibrium stock market macroscopic  state, for which $M=0$.

At the same time, in the econophysics world of the stock markets, it is much better 
to consider that an individual trader has three possible actions (or "spin states"):
"buying", "selling" or "waiting". (The transformation from one of these states to
another one is furthermore a discontinuous process due to a threshold being exceeded, 
usually, the price of stock).

It is evident, for us, that it is more adequate to introduce the Ising-1 spin model,
which involves three possible "spin states": $S_i\in \{+1,0,-1\}$.

As a matter of fact, the individual traders only have information on the action
of a limited number of other traders, and, generally, they can only  see the cooperative 
response market as a whole, in terms of an increase or decrease in the market
value.

In conventional economics, markets are assumed to be efficient if all available
information is reflected in current market prices. This constitutes a first order 
approach being based on the theory of "rational expectation"[3],[4]. 

If all the traders (in fact the "noise traders")
minimized their disagreement, i.e., if all the investors make
decision only depending on what other investors are doing, the market will end up in
one of two possible states, where all the traders will either  buy or  sell.
However, this does never happen in any real financial markets, because the traders never
can follow entirely the market opinion. 

In our construction it is 
possible to incorporate the existent different opinions ( which, from our point of view,
are not of random influence) considering the existence of more than one  
ordering mechanism, involving, macroscopically, more than one order parameter.

There are, indeed, many real systems for which the phase transition behavior cannot be completely
described by a singleorder parameter, because there are two (or more) types of ordering processes which
can take place. Such systems must be described, at least, by two (or more) order parameters, 
and their particular behavior lies in the mutual interaction between the two (or more) ordering processes.

From  microscopic point of view, we consider that there are both the long and short range interactions responsible for 
the phase transition behavior of these systems, but, is important to observe that
there are other effects such as the particularities of the constituents or the external field 
effects which can completely change the critical behavior.

In the following, we assume,  that, for  the  macroscopic behavior,
a good description of the critical financial
market phenomena cannot
be completely specified by a singleorder parameter
and we define two kinds of ordering correlations, which will be 
designated  as "orientational" and "structural" ordering, so, we  
identify, at least, two order parameters. 
For this purpose we consider, [5], the general Ising-like Hamiltonian
\begin{equation}
{\cal H}=-H\sum _i S_{i}-J\sum _{i,j}S_{i}S_{j}-D\sum _i (S_i)^2-K\sum _{i,j}(S_i)^2(S_j)^2
\end{equation}
which involves two mechanisms of ordering
(it was assumed that $K$ favors structural order and $J$ determines the 
orientational one),
with the two macroscopic order parameters, defined as  
\begin{equation}
M=<S_i>
\end{equation}
and
\begin{equation}
Q=<S_i^2>
\end{equation}
Obviously, these two parameters are not independent in the sense that, generally speaking,
$M\neq 0$ implies $Q\neq 0$.

So far, we considered the existence of one type of traders namely 
"noise traders" which doesn't develop their own opinion. In reality, the model
should be extended to take into account the action of "fundamentalist traders"[4]
which are the traders who makes their opinion on the stock of a company based on the analysis of the 
fundamentals of the company.
This opinion could be optimistic or pessimistic. It is possible that different
fundamentalist traders (trading agencies) can have different evaluations of the
fundamental value.
It is also possible that "fundamentalist traders" reverse their bias optimistic
or pessimistic when the price variation is to high or to low.
It is possible that the fundamentalist action to be qualitatively different of 
the rest of traders which are dominated by less rational behavior.

In order to take into account the opinion of both fundamentalist and noise 
traders, in the spin model we must consider 4 spin states: 
$S_i=\{-\frac{3}{2},-\frac{1}{2},\frac{1}{2},\frac{3}{2}\}$ corresponding
to the "buy" and "sell" options of both "noise traders" and 
"fundamentalist traders". In this case the "wait" option is not of any 
interest. By introducing this four spin states, we can use the 
Ising-$\frac{3}{2}$ spin model.

As a matter of fact, the market perception is realized at a macroscopic level,
so, we consider that the mean-field solutions for $M$ and $Q$ have the fundamental
importance in analyzing the critical features of the financial markets. At the same time
we consider that it is possible to influence this critical behavior from the 
analysis of the non-analyticity of $Q=Q(M)$ dependence.

In the mean-field approximation, solutions for $M$ and $Q$ were determined  in both
Ising $S=1$ and $S=\frac{3}{2}$ cases. (The last model in fact introduces four types of
different options for market investors, i.e., there are 4 "spin states":
$s_i\in \{\frac{3}{2},\frac{1}{2},-\frac{1}{2},-\frac{3}{2}\}$).

At the first stage, considering $H=0$, from our analysis,
it was observed, [6], that, as a consequence of the mathematical (kinematical) relationship
between Q and M, the only relevant $J$-values are in the range $0.55 < J < 7.5$.
It is because, for $J < 0.55$, the Q-order parameter values less than unity are not founded,
and, on the other hand, for $J > 7.5$, the curves are $J$- insensitive,
being almost of the same form. 

Following the results of [7], realized on spin-$1$ Ising Hamiltonian with 
renormalization group approach, it was evident that there is one fixed point 
of interest,
for which $J^*=0.5822, K^*=0.9944, D^*=-4.2449$.
It was very encouraging for us that these  values agrees with those obtained from
our analysis of the behavior of the both ordering parameters[6].

For both $S=1$ and $S=\frac{3}{2}$ models, it was observed
that, as a function of 
$\frac{D}{K}$ and $\frac{H}{J}$, 
there are possible bipolar-, 
quadrupolar-type and successive transitions. 

On the other hand, our analysis revealed that, in both cases, the $Q=Q(M)$ dependence 
is also strongly determined by
$\frac{D}{K}$ and $\frac{H}{J}$ ratios, and, in some cases, we obtained two values 
for the order parameter $Q$, both corresponding to the same value of $M$ [12]. 

If it is possible to introduce an external field control, via $D$ and $H$ couplings,
it is possible to move or even eliminate the existent non-analyticity.

We also consider that the expected phase transition results in a distinct abnormal 
mathematical dependence of the coupled positional and orientational order parameters.

In order to examine the analytical dependence of $Q=Q(M)$ we write the equation 
in the following form:
\begin{equation}
   f(Q,M)=Q-Q(M)=0
\end{equation}
We start by choosing values for $Q$ between $0$ and $1$, starting from 
$0$ with small $dQ$ increments. For every $Q$ we solve the above equation
and get the corresponding values for $M$.
We developed a computer program which solves the equation for different values 
of parameters[8].

It is interesting to see that, in the domain of interest for spin-1 model, 
the two order parameters have the same direction of variation (both are growing),
but in the $\frac{3}{2}$ Ising spin model we have a strong nonanalytical behavior.

\begin{figure}[htbp]
\input{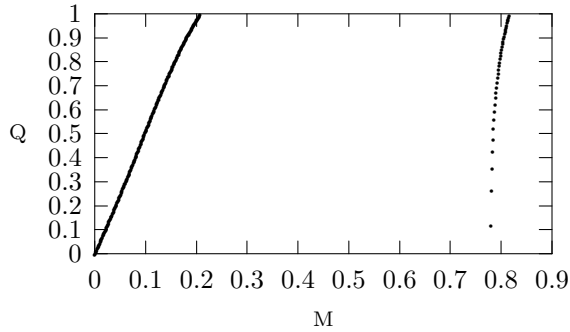}
\caption{$Q=Q(M)$ for $K=0.9944, J=0.5822, D=-4.2449, H=-1.6$}
\label{fig.1}
\end{figure}

In Figure~\ref{fig.1} we show the dependence $Q=Q(M)$ in Ising-$\frac{3}{2}$
model for a certain set of values of coupling constants corresponding to the fixed point
of the Ising-$\frac{1}{2}$ model.

\begin{figure}[htbp]
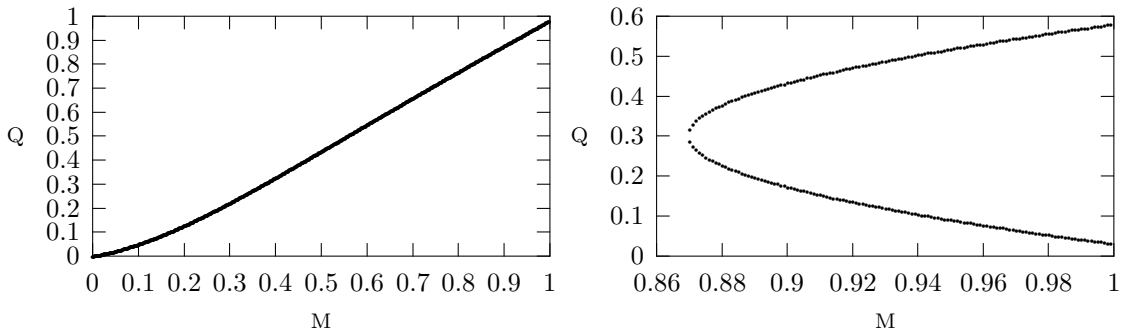

\parbox{7cm}{\input{FIG2A.TEX}}\parbox{7cm}{\input{FIG2B.TEX}}
\caption{$Q=Q(M)$ for $K=1, J=1, D=-1, H=0$ (spin 1 left, spin 3/2 right).}
\label{fig.2}
\end{figure}

\begin{figure}[htbp]
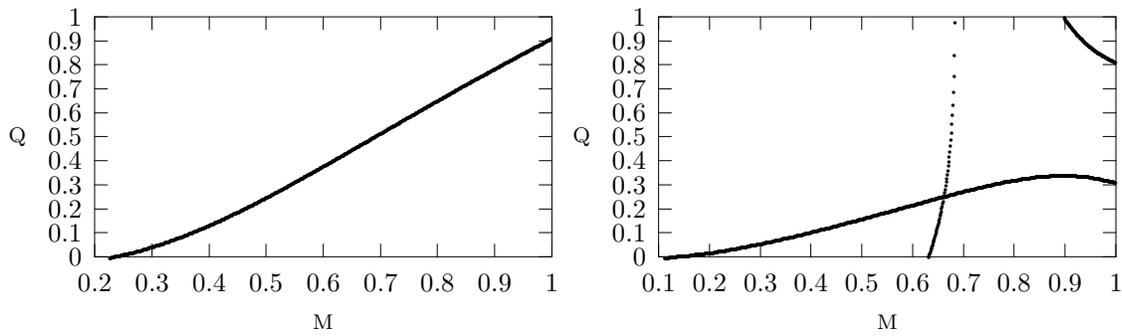

\parbox{7cm}{\input{FIG3A.TEX}}\parbox{7cm}{\input{FIG3B.TEX}}
\caption{$Q=Q(M)$ for $K=1, J=1, D=-0.5, H=-1.6$ (spin 1 left, spin 3/2 right).}
\label{fig.3}
\end{figure}

In Figures~\ref{fig.2} and \ref{fig.3} we present the dependence $Q=Q(M)$ 
in the two models for different meaningful values of coupling constants.

It's valuable the fact that for the same values of the cuplling constants
(K,J), the two ordering fields D and H modify in a different manner
the analyticity of $Q=Q(M)$.

\end{document}